\documentclass[aps,showpacs,preprintnumbers,amsmath,amssymb, 
twocolumn, tightenlines,
]{revtex4}

\usepackage[dvips]{graphicx}
\usepackage{bm}

\begin{document}

\bibliographystyle{apsrev} 

\title {Multi-fermion states for heavy fermions bound via Higgs exchange}

\author{M.Yu.Kuchiev} \email[Email:]{kmy@phys.unsw.edu.au}

\affiliation{School of Physics, University of New South Wales, Sydney
  2052, Australia}

%

    \date{\today}

    \begin{abstract} 
A possibility to produce bound states of several heavy fermions, which are bound together due to the Higgs exchange, is examined. It is shown that for 12 fermions, 6 fermions and 6 antifermions, occupying the lowest $S_{1/2}$ shell this bound state cannot exist if the fermion mass is below the critical value $m_{\mathrm cr}$, which depends on the Higgs mass and is found to be restricted to $320 < m_{\mathrm cr} < 410$ Gev$/c^2$ for the Higgs mass $m_{\mathrm H}$  in the interval $100 \le m_{\mathrm H}\le 200$ Gev$/c^2$. This estimate is derived in the relativistic mean field approximation. The corrections are estimated to be not able to reduce significantly the critical value for the fermion mass. Consequently there exist no bound state for 12 top quarks, and the only feasible hope to observe a bag of 12 fermions experimentally should rely on possible existence of heavy fermions of the next, fourth generations.
    \end{abstract}


    \maketitle



The Standard Model explains the mass spectrum of fermions via their interaction with the Higgs field. The heavier the fermion is the stronger is its interaction with the Higgs field. Since the Higgs is a scalar this interaction has an attractive nature, which makes it tempting to look for new phenomena related to the heaviest known fermion, the top quark, as well as consider what may happen if heavier fermions of the next (fourth) generation exist in nature. It was suggested previously by Froggatt et al \cite{Froggatt:etal} that a system of 12 top quarks, 6 top and 6 anti-top occupying the lowest $S_{1/2}$ orbital can be bound by the forces originating from the Higgs exchange. This interesting suggestion was based on a simplified description, which neglected the Higgs mass. In addition, it was supported by estimations based on the Hydrogen-atom model. However, our more accurate calculations based on the nonrelativistic mean field  approach \cite{Kuchiev:2008fd}, which were supported by \cite{Richard:2008uq}, indicated that the binding is nonzero, albeit very small, only in the mentioned simplified approach, which neglects the Higgs mass. An account of the Higgs mass, which is limited by the current experimental bounds,  makes this 12 top-antitop system unbound. Recently Froggatt and Nielsen \cite{Froggatt:2008hc} discussed the issue anew.  This latter work claims again that the binding state of 12 top quarks exists and, moreover, the binding is very strong. This claim was supported by model-type approaches, which present a modification of the ones used previously in \cite{Froggatt:etal}.  

Having in mind to clarify the issue this papers examines the fermion bound state for 12 fermions occupying $S_{1/2}$ shell considering the mass of the heavy fermion $m$ as a parameter. The  aim is  to find a critical value for this mass $m_{\mathrm cr}$ below which, when $m\le m_{\mathrm cr}$, the mentioned bound state cannot exist. The mean field approximation for the relativistic system that contains fermions interaction with the Higgs field is employed. The critical mass found depends strongly on the Higgs mass $m_{\mathrm H}$. 
Taking the later in the interval $100\le m_{\mathrm H}\le 200$ Gev$/c^2$ we find the following restriction on the critical fermion mass $320 < m_{\mathrm cr}< 410$ Gev$/c^2$. 

Consider $N$ heavy fermions that interact with the Higgs field $\Phi$. Take  the conventional unitary gauge where $\Phi$ is represented by the real field $ \xi$
\begin{equation}
\Phi\,=\,\frac{v}{\sqrt{2}}\left( \begin{array}{c} 0 \\ \xi \end{array}\right)~.	
\label{Phi-xi}	
\end{equation}
Here $v=246$ Gev is the VEV, which is achieved when $\xi=1$. The part of the Standard Model Lagrangian, which describes the interacting Higgs $\xi$ and fermion fields $\psi$ reads ($\hbar=c=1$)
\begin{align}
\!{\cal L}=
\frac{v^2}{2}\Big(\partial^\mu {\xi} \partial_\mu {\xi}\!-\!\frac{m_{\mathrm H}^2}{4}(\xi^2\!-\!1)^2\Big)\!+
\bar \psi( i\gamma^\mu \partial_\mu \!-m \xi ) \psi\,.
\label{L}
\end{align}
The important feature of the problem is that the fermion mass $m$ is presumed large, which allows us to neglect in the first approximation interactions with all other fields, gluons, $Z^{\,0}$ and $W$ mesons etc, which are omitted in (\ref{L}).
Moreover, the large fermion mass and the large number of fermions considered, $N=12$, allow one to rely in the lowest approximation on the mean field approach. Using it we replace the fields in Eq.(\ref{L}) by stationary wave functions calling them $\xi$ for the Higgs field and $\psi$ for the single-particle wave functions of fermions. Searching for the spherically symmetrical solution, take $\xi=\xi(r)$ and assume that all fermions occupy the same shell with total and angular momenta $j$ and $l$ (ultimately we will concentrate on the $S_{1/2}$ state). In this state the fermions are described by the same large $F(r)$ and small $G(r)$ components of the Dirac spinor. Using Eq.(\ref{L}) one writes the Hamiltonian $H$ of the system
\begin{align}
H\,=\,&\int_0^\infty\,
\Big[ \,\,\frac{v^2}{2}\,\Big(\,\xi'^2+\frac{1}{4}\,m_{\mathrm H}^2\,(\xi^2-1)^2
\,\Big)\,(4 \pi\,r^2) 
\label{HFG}
\\
& + N\,\Big(\,2\big(F'+\frac{\varkappa}{r} F\,\big)\,G+m \xi\, (\,F^2-\,G^2\,)\,\Big)\,\Big]\,dr~.
\nonumber
\end{align}
Here $\varkappa=\pm\,(j+1/2)$ for $l=j\,\pm 1/2$, and normalization $\int_0^\infty(F^2+G^2)\,dr=1$ is taken. 
From Eq.(\ref{HFG})  one derives the 
mean-field equations for $\xi,F,G$
\begin{align}
\xi''+\frac{2}{r}\,\xi'&+\frac{m_{\mathrm H}^2}{2}\,\,\xi\,(1-\xi^2)\,=\,\frac{(N-1)m}{4\pi \,v^2}\,\frac{F^2-G^2}{r^2}~,
\label{xi}
\\
&(\varepsilon-m\,\xi)\,F~\,=\,-G'\,+\,(\varkappa/r)\,G~,
\label{F}
\\
&(\varepsilon+m\,\xi)\,G~\,=\,\,~~F'\,+\,(\varkappa/r)\,F~,
\label{G}
\end{align}
where the eigenvalue $\varepsilon$ is positive, and the factor $N-1$ in Eq.(\ref{xi}) takes into account the fact that each fermion interacts only with $N-1$ other fermions. These equations are supplemented by the boundary conditions, which for the localized fermion wave function have a conventional form
\begin{equation}
F(0)=F(\infty)=G(0)=G(\infty)=0~.
\label{FFGG}
\end{equation}
For the Higgs field the boundary conditions read
\begin{equation}
\xi'(0)\,=\,0~,\quad \quad \xi(\infty)\,=\,1~.
\label{xi0}
\end{equation}
The second one guarantees that outside the fermion bag the Higgs field takes its VEV, 
while the first one makes the Higgs field regular at the origin by suppressing a singularity 
at $r=0$ in the term $(2/r) \xi'$ in Eq.(\ref{xi}).
\begin{figure}[h]
\centering
\includegraphics[height=5.5 cm,keepaspectratio = true, 
]{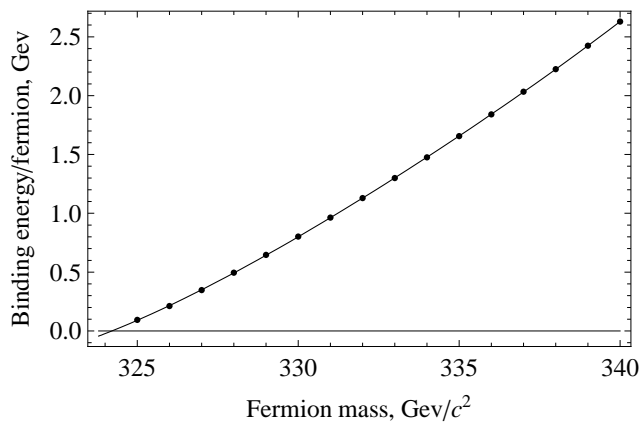}
\caption{ \label{ras} 
Binding energy per fermion derived from the total energy $H$ in Eq.(\ref{HFG}) versus the mass $m$ of heavy fermions when 12 fermions
(6 fermions and 6 antifermions) occupy the same $S_{1/2}$ shell. The Higgs mass is $m_\text{H}=100$ Gev$/c^2$. The dots present calculations based on Eqs.(\ref{HFG})-(\ref{G}). An interpolating solid line shows that the bound state exists only for $m\ge m_{\mathrm cr}=324.2$ Gev$/c^2$.}
\end{figure}

The behaviour of the system described by Eqs. (\ref{HFG}) - (\ref{G}) was studied in Ref. \cite{Kuchiev:2008gt} for large fermion masses $m>1$ Tev$/c^2$, when the ultrarelativistic description is valid. It was found that the system reacts to the presence of the large fermion mass in such a way as to eradicate its influence on its physical parameters. In particular, the mass of the bag proves be independent on $m$. This phenomenon was described both by numerical methods and by the semi-analytical solution that is valid in the ultrarelativistic region.
Ref. \cite{Kuchiev:2008fd} studied the case of top quarks, when the nonrelativistic description is valid. In this situation the Higgs field proves be close to its VEV, $\xi\approx 1$, and consequently the linearized in $\xi$ version of Eq.(\ref{xi}) is applicable, which allows one to resolve Eq.(\ref{xi}) reducing the system (\ref{xi})-(\ref{G}) to a single equation on the nonrelativistic wave function of the fermion.

Here we consider the case when $m$ can be larger than the top quark mass, but remains below the ultrarelativistic limit,  $170<m<1000$ Gev$/c^2$. In this region the nonrelativistic approximation can become unreliable, while the mentioned analytical approach for the ultrarelativistic case is also not applicable. We are therefore to rely on  numerical methods solving Eqs.(\ref{xi})-(\ref{G}) directly, without additional simplifications. Fig. \ref{ras} shows the results of these calculations presenting the binding energy $E_{\text B}$ of the bag of $N=12$ fermions (6 fermions and 6 antifermions), which occupy the same $S_{1/2}$ shell. The binding energy per fermion shown in Fig. \ref{ras} equals $E_{\text B}/N =m-H/N$ where $H$ is the total energy (\ref{HFG}) calculated with the wave functions found from solution of Eqs. (\ref{xi})-(\ref{G}).
Fig. \ref{ras} presents this energy versus the mass of the fermion $m$. The positive binding indicates that the bound state is stable. From data presented in Fig. \ref{ras} we derive that the bag can exists provided $m\ge m_{\mathrm cr}\simeq 324.2$ Gev$/c^2$.
\begin{figure}[t]
\centering
\includegraphics[height=5.7 cm,keepaspectratio = true, 
]{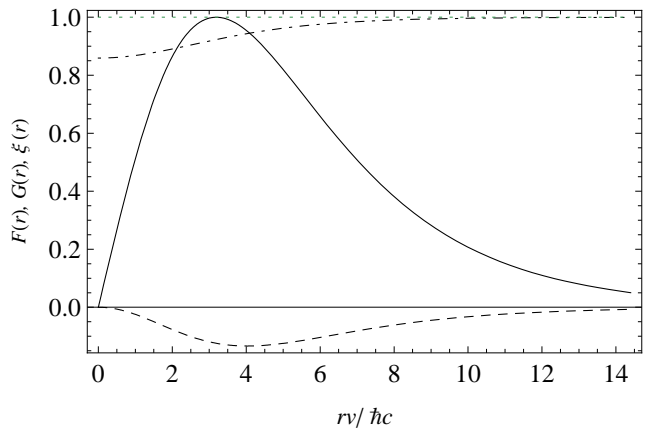}
\caption{
 \label{dva} Fields inside the fermion bag. 
The Higgs and fermion masses are $m_\text{H}=100$, $m=328$ Gev$/c^2$. 
Solid and dashed lines - large $F(r)$ and small $G(r)$ components of the Dirac spinor (scaled to make $\max(F)=1$), dot-dashed line - Higgs field $\xi(r)$, all solutions of Eqs.(\ref{xi})-\ref{G}), dotted line - nonperturbed Higgs $\xi_0=1$. The distance from the center of the bag $r$ is scaled by the VEV of the Higgs field $v=246$ Gev.}\end{figure}
\noindent

The wave functions $F(r)$, $G(r)$ and $\xi(r)$, which describe 
the fermions and Higgs field inside the bag, are shown in Fig. \ref{dva}, which presents them for $m_\text{H}=100$, and the fermions mass $m=328$ Gev$/c^2$. This behaviour of the wave functions is typical when fermion mass just above the critical value. 
In this region the Higgs field is only slightly modified inside the bag, while the large component of the fermion radial wave function $f(r)=F(r)/r$ is prominent at the center of the bag and smoothly decreases in the outside region. It is instructive to compare this behaviour with the case when the fermion is very heavy, $m>1$ Tev, which was considered in \cite{Kuchiev:2008gt}. In this ultrarelativistic case the Higgs field develops a node on the surface of the bag becoming negative inside, while the fermion wave function is localized on the surface of the bag, in the vicinity of the node of the Higgs field. Thus, for very heavy fermions the fermion wave function is suppressed at the center of the bag, the bag can be called `hollow'. In contrast, in the considered presently case of large, though moderately large fermion masses, the fermion wave function $f(r)$ is large at the origin, the bag is `solid'. Observe that Fig. \ref{dva} shows that the bag has no sharp, well defined surface, which contrasts the nuclear structure. This property complies with the fact that the range of forces between fermions $r\sim 1/m_{\mathrm H}$ induced by the Higgs exchange inside the bag is comparable with the size of the bag, while nuclear forces have a short-range nature in comparison to the size of large nuclei.

\begin{figure}[t]
\centering
\includegraphics[height=5.5 cm,keepaspectratio = true, 
]{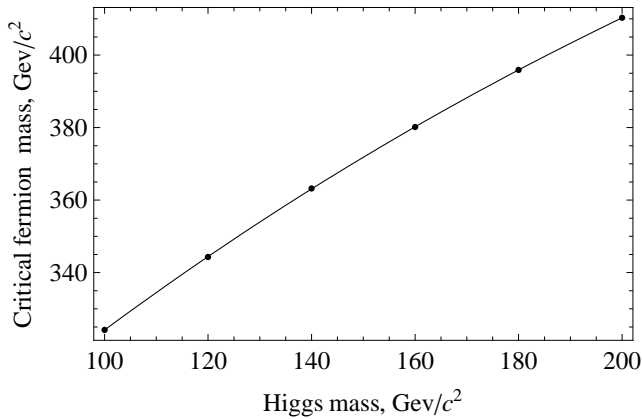}
\caption{
 \label{tri} The critical fermion mass $m_{\mathrm cr}$ versus the Higgs mass $m_{\mathrm H}$. For heavy fermions with $m>m_{\mathrm cr}$ there can exist a bag of 12 fermions 
occupying $S_{1/2}$ shell, which are bound via the Higgs field. 
The dots present data from numerical solution of Eqs.(\ref{xi})-(\ref{G}), the solid line is an interpolation.}
\end{figure}

The latter comment reminds us that the  fermion binding energy may depend on the Higgs mass. To study this dependence the calculations were repeated for different values of $m_{\mathrm H}$ in the most interesting region $100\le m_{\mathrm H}\le 200$ Gev$/c^2$. For each $m_{\mathrm H}$ from this region a set of the fermion masses $m$ was taken and the critical fermion mass $m_{\mathrm cr}$ was found using a procedure illustrated by Fig. \ref{ras}. The found from these calculations critical fermion mass is plotted in Fig. \ref{tri} versus the Higgs mass. We see that the critical mass quickly rises with the Higgs mass, which complies with the fact that the heavy Higgs reduces the range in which an effective attraction between fermions exists. From data plotted i Fig. \ref{tri} we derive that if the Higgs mass is restricted by the interval $100\le m_{\mathrm H} \le 200$ Gev$/c^2$ then the critical fermion mass lies in the interval  $320 <m_\mathrm{cr}<410$ Gev$/c^2$.


\noindent

Assessing the accuracy of these results one should keep in mind errors related to several phenomena neglected in the approximation considered. An important class among them represent radiative corrections, which importance was emphasized previously in
\cite{Dimopoulos:1990at,Bagger:1991pg,Farhi:1998vx,Crichigno:2009kk}. 
It was alleged that these corrections are able to make the bag shallow 
\cite{Dimopoulos:1990at,Crichigno:2009kk},  deflate it into inexistence \cite{Farhi:1998vx}, or ultimately even make the vacuum unstable \cite{Bagger:1991pg}. It seems that the issue is important and need further consideration; we are planning to address it in some detail in the near future. Presently, one can state that the reported results \cite{Dimopoulos:1990at,Bagger:1991pg,Farhi:1998vx,Crichigno:2009kk} indicate on a repulsive nature of radiative corrections. Therefore in any case they cannot reduce the estimates for the critical mass of heavy fermions.

There exist a number of other corrections. One class of them represent the many-body corrections, which include deviation of the many-body wave function from the mean field approximation, as well as the retardation and recoil corrections. One should anticipate though that the considered system of large number of tightly bound fermions, 12 fermions located in the region of the size of $r\sim \hbar c/v$, $v=246$ Gev$/c^2$, makes the mean field approach sensible, many body corrections should not be very pronounced. There exit also corrections that arise due to interactions with gluons, $Z^0$ and $W$ bosons etc, but for the considered prominent fermion masses $>300$ Gev$/c^2$ these interactions are certainly much weaker than the one induced by the Higgs exchange. Speaking of which one could remember that there exists also an interaction between fermions, which is induced by the virtual fermion-antifermion annihilation into the Higgs. However, for the considered heavy fermions this channel is strongly suppressed due to the fact that the virtual Higgs, which arises from an annihilation, propagates with large virtual frequency $\omega\simeq 2m$, which greatly reduces contribution of this process(similar situation reveals itself in the positronium, where the annihilation channel provides only a small correction.)

This discussion shows that there is no reason to unticipate that any of the mentioned phenomena would be able to change the main qualitative conclusion of this work.  The critical mass for fermions below which they are certainly not able to produce a bag (12 fermions in $S_{1/2}$ shell bound by the Higgs exchange) is large, $m>320$ Gev$/c^2$. This bound greatly exceeds the mass of the top quark.  Consequently 12 top quarks are not able to form the bound state. This important qualitative conclusion complies with previous calculations \cite{Kuchiev:2008fd,Richard:2008uq} and contradicts results of Froggatt and Nielsen \cite{Froggatt:2008hc}. In order to find a bridge  between conclusions of \cite{Kuchiev:2008fd,Richard:2008uq} and \cite{Froggatt:2008hc}
Froggatt and Nielsen put forward an idea that there may exist a particular phase transition, a phenomenon which manifests itself as an existence of two different states, which have the same quantum numbers. These authors alleged that one of these states, which is bound for top quarks, is studied in \cite{Froggatt:2008hc}, while another one, which remains unbound, presents itself in calculations of \cite{Kuchiev:2008fd,Richard:2008uq} and, developing this argument, in the present work. This idea is supported in \cite{Froggatt:2008hc} by arguments derived from the proposed in this work `toy model', which makes a number of simplifications.

In order to check the idea that there may exist some other bound state, in which 
the Higgs field is also spherically symmetrical and the fermion wave function has the same $S_{1/2}$ symmetry, the following numerical approach was used. A set of initial approximations for the Higgs field was taken. This set included, in particular, the wave functions for the Higgs field, which described it as being mostly flat inside the bag, in accord with presumptions of the `toy model' of \cite{Froggatt:2008hc}. Each initial approximation was used as input data for a numerical procedure, which generated the self-consistent numerical solution of the mean field equations. It was found that for any sufficiently  smooth initial approximation the numerical procedure converges, and converges to one and only self-consistent solution, regardless of the initial approximation. Thus, there was found no indications of a possible existence of a different bound state with the given symmetry. Summarizing, we have a situation when direct numerical calculations of this work, as well as of \cite{Kuchiev:2008fd,Richard:2008uq} show that top quarks are too light to be able to produce the bound state in question.

In conclusion, we investigated a possibility that 12 heavy fermions, which occupy the same $S_{1/2}$ shell can be bound by an attraction, which originates from the Higgs field exchange. The fermions in question need to be heavy to make this attraction sufficiently strong. The critical mass for fermions, below which this bound state cannot exist, is found using the relativistic mean field approximation. A prominent dependence of this critical mass on the mass of the Higgs field is found. Taking the later in the interval $100\le m_{\mathrm H}\le 200 $ Gev$/c^2$ the critical fermion mass is found in the interval $320\le m_{\mathrm H}\le 410 $ Gev$/c^2$. The fact that even the lowest estimate from this set is almost twice as heavy as the mass  of top quarks makes it unrealistic to expect that top quarks are able to produce the mentioned bound state. The only hope to observe experimentally the multi-fermion bound states, which exists due to the Higgs exchange, could rest on a hope that there exist heavy fermions of the next, fourth generation.

This work is supported by the Australian Research Council

\end{document}